\documentclass[11pt]{article}
\newcommand{\be}{\begin{equation}}
\newcommand{\ee}{\end{equation}}
\newcommand{\beq}{\begin{eqnarray}}
\newcommand{\eeq}{\end{eqnarray}}

\def\theequation{\arabic{section}.\arabic{equation}}
\usepackage{graphicx}
\begin{document}
%\pagestyle{empty}
%\begin{flushright}
%{BROWN-HET-1117} \\
%{March 1998}
%\end{flushright}
\begin{center}
{\bf \LARGE Non trivial generalizations of the Schwinger pair production result II}\\
[7mm]
H. M. FRIED%\footnote{Supported in part by a CNRS/Brown Accord }
\\
{\em Department of Physics \\
Brown University \\
Providence R.I. 02912 USA}\\
[5mm]
Y. GABELLINI%\footnote{Supported in part by a CNRS/Brown Accord }
\\
{\em Institut Non Lin\'eaire de Nice\\
UMR 6618 CNRS\\
 1361 Route des Lucioles\\
06560 Valbonne France}\\
[5mm]

\vspace{5mm}
Abstract
\end{center}
It is suggested that Schwinger's (1951) vacuum persistence probability against pair production by an intense but constant electric field is a very good approximation to the corresponding quantity if the field does not vary aprreciably over distances less than $m/e|E|$.
\newpage
%{\bf\section{Introduction}}
\def\theequation{\arabic{equation}}
\setcounter{equation}{0}
The object of this brief letter is to point out an argument previously overlooked, which suggests that the vacuum persistance probability for the production of charged lepton pairs in Schwinger' seminal 1951 paper \cite{one}, calculated for constant electromagnetic fields, is, in fact, a very good approximation to the same quantity in the limit of sufficiently intense fields. Previous approximations \cite{two} have given qualitative results for `` slowly--varying '' fields; but the relevant observation here is that, unless those fields vary appreciably over space--time distances on the order of $m^{-1}(m^2/e\vert F\vert)$, which for intense fields can be far smaller than the Compton wavelenght of each member of the produced pair, a respectable approximation to Schwinger's formula is obtained by merely replacing his constant $F_{\mu\nu}$ by $F_{\mu\nu}(x)$. 

For equivalently small distances, we also note that the vacuum persistance probability for the production of high mass charged tachyon pairs, as suggested in a recent QTD model \cite{three}, can be given an expression quite similar to that of Schwinger; and that the resulting approximation for intense, external fields is essentially the same. Estimates of lepton pair production by crossed laser beams of high intensity \cite{four} are also simplified.

We follow the notation and functional techniques previously used \cite{two,five} for this problem, in terms of a convenient Fradkin representation \cite{six}, where the vacuum persistance probability $P$ is given by $P=\exp[2 {\rm Re} L[A]]$, with :
\beq
{}&L[A] = -\displaystyle{1\over 2}\int_0^{\infty}\!{ds\over s}\,e\,^{\displaystyle{-ism^2}}\!\!\exp\,\biggl[i\int_0^s\!ds'\sum_{\mu}{\delta^2\over\delta v_{\mu}^2(s')}\biggr]\nonumber\\
&\times\displaystyle\int\!{d^4p\over(2\pi)^4}\,\,e\,^{\displaystyle{ip\!\cdot\!\!\!\int_0^s\!ds'v(s')}}\displaystyle  \int \!d^4\!x\,e\,^{\displaystyle -ie\int_0^s\!ds'v_{\mu}(s') \,A_{\mu}(x-\int_0^{s'}\!\!v)}\\
&\times\,{\rm tr}\biggl(\displaystyle e\,^{\displaystyle e\!\int_0^s\!ds'\sigma\!\cdot\!F(x-\int_0^{s'}\!\!v)}\biggr)_+\Bigg\vert_{\,v_{\mu}\rightarrow 0}\nonumber
\eeq
In fact, $L[A]$ is a completely gauge invariant functional of the $F_{\mu\nu}$; this may be seen by following the little known functional argument presented in the book of ref\cite{six}, wherein the $A_{\mu}(x-\int_0^{s'}\!\!v)$ of (1) may, for this closed--loop calculation, be replaced by :
\be
A_{\mu}(s')\equiv A_{\mu}(x,u(s'))\equiv\int_0^1\!\!\lambda\,d\lambda \,F_{\mu\nu}(x-\lambda u(s'))u_{\nu}(s') 
\ee
with $u_{\nu}(s')=\int_0^{s'}\!\!ds''\,v_{\nu}(s'')$; and the $L[A]$ of (1) is then clearly independent of gauge : $L[A]=L[F]$.

As in \cite{two}, it is convenient to formally introduce $u_{\mu}(s')$ as an independent variable, by means of the functional identity : 
$$1=\int\!\!d[u]\,\delta[u(s')-\int_0^{s'}\!\!v]=N'\int\!\!d[u]\,\int\!\!d[\Omega]\,e\,^{\displaystyle i\int_0^s\!ds'\Omega_{\mu}(s')[u(s')-\int_0^{s'}\!\!v]}$$
where $N'$ is the appropriate normalization constant. With the aid of (2), (1) can be rewritten as : 
\beq
{}&L[F] = -\displaystyle{1\over 2}\int_0^{\infty}\!{ds\over s}\,e\,^{\displaystyle{-ism^2}}N'\!\!\int\!\!d[u]\int\!\!d[\Omega]\,e\,^{\displaystyle i\!\!\int\!\!\Omega u}\!\!\int\!\!{d^4p\over(2\pi)^4} \int \!\!d^4\!x\,\,{\rm tr}\biggl(\displaystyle e\,^{\displaystyle e\!\int_0^s\!ds'\sigma\!\cdot\!F(x-u(s'))}\,\biggr)_+\nonumber\\
&\times\,e\,^{\displaystyle i\!\!\int_0^s\!{\delta^2\over\delta v^2}}e\,^{\displaystyle i\int_0^s\!ds'v_{\mu}(s') \,[p_{\mu}-eA_{\mu}(s')-\int_{s'}^{s}\!\!\Omega(s'')]}\Bigg\vert_{\,v_{\mu}\rightarrow 0}
\eeq
where Abel's trick has been used in rewriting $\int_{0}^{s}\!ds' \Omega_{\mu}(s')\int_{0}^{s'}\!\!ds'' v_{\mu}(s'')$. 

The functional linkage operation of (3) is immediate, and one finds for the second line of (3) : 
\beq
{}&\hskip-0.3truecm\exp\Bigl[{\displaystyle -i\!\!\int_0^s\!\!ds' \,[p-eA(s')-\!\int_{s'}^{s}\!\!\Omega(s'')]^2\Bigr]}=\exp\Bigl[\displaystyle-isp^2 -ie^2\!\!\int_{0}^{s}\!\!A^2- i\!\!\int_{0}^{s}\!\!\!\int_{0}^{s}\!\!ds_1ds_2\,\Omega_{\mu}(s_1)h(s_1,s_2)\Omega_{\mu}(s_2)\nonumber\\
&\displaystyle+2iep\!\cdot\!\!\!\int_0^s\!\!A+2ip\!\cdot\!\!\!\int_0^s\!\!ds's'\Omega(s')-2ie\!\!\int_0^s\!\!ds'\,\Omega_{\mu}(s')\!\!\int_0^{s'}\!\!ds''A_{\mu}(s'')\Bigr]\eeq
where $h(s_1,s_2)=\!\int_0^s\!ds'\,\theta(s'-s_1)\theta(s'-s_2)=\theta(s_1-s_2)s_2+\theta(s_2-s_1)s_1$. The $p$--dependent terms of (4) then yield :
\be\int\!\!{d^4p\over(2\pi)^4}\,\,e\,^{\displaystyle-isp^2+2ip\!\cdot\!\!\int_0^s\!\!ds' \,[eA(s')+s'\Omega(s')]}={-i\over16\pi^2s^2}\exp\left\{{i\over s}\left[\int_{0}^{s}\!\!ds'\Bigl( eA(s') + s'\Omega(s')\Bigr)\right]^2\right\}\ee
and, combining (5) with the remaining of (3), one finds :
\beq
{}&L[F] = \displaystyle{i\over 32\pi^2}\int_0^{\infty}\!{ds\over s^3}\,e\,^{\displaystyle{-ism^2}} \int \!\!d^4\!x\,N'\!\!\int\!\!d[u]\int\!\!d[\Omega]\,{\rm tr}\biggl(\displaystyle e\,^{\displaystyle e\!\int_0^s\!ds'\sigma\!\cdot\!F(x-u(s'))}\,\biggr)_+\nonumber\\
&\times\,e\,^{\displaystyle-ie^2\Bigl[\int_0^s\!\!ds' \,A^2(s')-{1\over s}\left(\int_0^s\!\!ds' \,A(s')\right)^2\,\Bigr]}\\
&\times\,e\,^{\displaystyle-i\int_{0}^{s}\!\!\!\int_{0}^{s}\!\!ds_1ds_2\,\Omega_{\mu}(s_1)J(s_1,s_2)\Omega_{\mu}(s_2)+i\int_{0}^{s}\!\!ds'f_{\mu}(s')\,\Omega_{\mu}(s')}\nonumber\eeq 
where $J(s_1,s_2)=h(s_1,s_2) - s_1s_2/s$, and $f_{\mu}(s')=u_{\mu}(s') - 2e\int_{0}^{s}\!\!ds'' A_{\mu}(s'')[\theta(s'-s'') - s'/s]$.

The observation we wish to make follows upon performing the Gaussian functional integral :
\be\sqrt{N'}\int\!\!d[\Omega]\,e\,^{\displaystyle-i\!\int\!\!\!\int_{0}^{s}\!\!\Omega_{\mu}J\Omega_{\mu}+i\!\int_{0}^{s}\!\!f_{\mu}\,\Omega_{\mu}}=e\,^{\displaystyle-{1\over 2}{\rm Tr}\ln(2J)}\,e\,^{\displaystyle+{i\over4}\int\!\!\!\int_{0}^{s}\!\!f_{\mu}J^{-1}f_{\mu}}\ee
where $J^{-1}(s_1,s_2)$ satisfies $\delta(s_1-s_2)=\int_0^s\!\!ds'J^{-1}(s_1,s')J(s',s_2)=\int_0^s\!\!ds'J(s_1,s')J^{-1}(s',s_2)$

The functional integral ( FI ) that results from the combination of (7) and (6) cannot, in general, be evaluated exactly; but, because it can be represented as a gaussian--weighted FI over $u_{\mu}(s')$, proportional to :
\beq{}&\displaystyle\sqrt{N'}\int\!\!d[u]\,e\,^{\displaystyle {i\over4}\!\int\!\!\!\int_{0}^{s}\!\!f_{\mu}J^{-1}f_{\mu}}\,{\rm tr}\biggl(\displaystyle e\,^{\displaystyle e\!\int_0^s\!ds'\sigma\!\cdot\!F(x-u(s'))}\,\biggr)_+\nonumber\\&\times\,e\,^{\displaystyle-ie^2\Bigl[\int_0^s\!\!ds' \,A^2(s')-{1\over s}\left(\int_0^s\!\!ds' \,A(s')\right)^2\,\Bigr]}\eeq
and contains non--gaussian corrections of $u$--dependence buried inside fields of form $F_{\mu\nu}(x-u(s'))$ and $\int_0^1\!\lambda\,d\lambda \,F_{\mu\nu}(x-\lambda u(s'))u_{\nu}(s')$, certain approximations are indeed possible.

Firstly, let us assume that $e\vert F\vert/m^2\le1$, $f_{\mu}(s')\sim u_{\mu}(s')$, i.e., the fields are weak. Then (8) becomes essentially a gaussian FI, with the natural constraints on the size of $\vert u(s')\vert$ given by $\vert\int\!\!\!\int uJ^{-1}u\vert\le1$. Since $O(J^{-1})\sim1/s^3$, and because the scale of $s$ is set by the factor $\exp[-ism^2]$, $O(sm^2)\le1$, it follows that $O(\vert u\vert)\le\sqrt s\le1/m$. Hence, if the fields $F_{\mu\nu}(x-\lambda u)$ do not vary appreciably over distances of a Compton wavelenght, the $u(s')$ dependence inside the $F_{\mu\nu}$ can be neglected; one then finds that the term of (8) proportional to $\exp\Bigl[-ie^2s<(A-<A>)^2>\bigr]$ becomes unity\footnote{This statement is literally true if only an electric field ${\cal E}_3(x_{\pm})$ in one spatial direction is considered, where $x_{\pm}=x_3\pm x_0$; but the inclusion of this exponential factor cannot change our order--of--magnitude estimates, unless exceptional cancellations were to occur.}, and one is left with the Schwinger, constant--field solution, although written in an unconventional way, where $F_{\mu\nu}\rightarrow F_{\mu\nu}(x)$. In comparison with Schwinger's constant electric field result for the vacuum persistence probability :
\be 
P=\exp\Bigl\{-{\alpha\over\pi^2}\int \!\!d^4x\,E^2\sum_{n=1}^{\infty}{1\over n^2}\,e\,^{\displaystyle-n\pi m^2/e\vert E\vert}\Bigr\}
\ee
The restriction $e\vert E\vert/m^2\le 1$ means that each term in the expansion of $\ln P$ is exponentially small, $P$ remains close to unity, and the probability of pair production is very small.
\vskip0.3truecm
In contrast, let us now assume that $e\vert E\vert/m^2>1$. Then the dominant term of $f_{\mu}(s')$ is not $u_{\mu}(s')$ but, rather, $-2e\!\int_0^s\!ds''[\theta(s'-s'')-s'/s]\!\int_0^1\!\lambda d\lambda \,F_{\mu\nu}(x-\lambda u(s'))u_{\nu}(s')$, which produces a quadratic gaussian weighting in (8) of order $e^2\vert F\vert^2\int\!\!\!\int uJ^{-1}u$, which quantity is now restricted to be $<1$. This, in turn, leads to the condition that $O(\vert u\vert)<{1\over m}\Bigl({m^2/e\vert F\vert}\bigr)$, which can be much smaller than the bound of the weak field case. Therefore, unless the intense fields vary significantly over distances less than $(m/e\vert F\vert)$, one is justified in dropping the $u(s')$ dependence inside the $F(x-\lambda u(s'))$ and $F(x-u(s'))$ of (8); when this is done, there is no difference between Schwinger's constant field solution for $P$ and the expression generated by (8), except that $F_{\mu\nu}\rightarrow F_{\mu\nu}(x)$. And, in this case, the exponential terms in the expansion of $\ln P$ of (9) can take on significant values, at least until $n=n_{max}\sim e\vert E\vert/\pi m^2$.
\vskip0.3truecm
For realistic external fields $F_{\mu\nu}(x)$, it is just that simple. This observation should have been made by the present authors two years ago, in ref\cite{two}.


\begin{thebibliography}{**}
\bibitem{two} J. Avan, H.M. Fried and Y. Gabellini, Phys. Rev. D{\bf67}, 16003 (2003). We take this opportunity to correct the misprint which occured in eq.(18) of this reference, where the exponential factors of $\delta/\delta u_{\mu}$ were incorrectly written in place of the correct $\delta/\delta f_{\mu}$, so that $\delta/\delta u_{\mu}$ should be replaced by $\int(\delta u_{\nu}/\delta f_{\mu})(\delta/\delta u_{\nu})$.
\end{thebibliography}
\end{document}